\documentclass[longauth,referee]{aa}
\usepackage{txfonts}
\usepackage{graphicx}
\usepackage{hyperref}

\begin{document}
\title{First light observations of the solar wind in the outer corona with the Metis coronagraph}

\author{M. Romoli\inst{\ref{unifi},\ref{ass}} \and
E.~Antonucci\inst{\ref{ass}} \and
V.~Andretta\inst{\ref{inaf-oac}} \and
G.~E.Capuano\inst{\ref{unict},\ref{inaf-oact}} \and
V.~Da~Deppo\inst{\ref{ifn-cnr},\ref{ass}} \and
Y.~De~Leo\inst{\ref{mps},\ref{unict}} \and
C.~Downs\inst{\ref{psi}} \and
S.~Fineschi\inst{\ref{inaf-oato}} \and
P.~Heinzel\inst{\ref{asu}} \and
F.~Landini\inst{\ref{inaf-oato}} \and
A.~Liberatore\inst{\ref{unito},\ref{inaf-oato}} \and
G.~Naletto\inst{\ref{unipd},\ref{ifn-cnr},\ref{ass}} \and
G.~Nicolini\inst{\ref{inaf-oato}} \and
M.~Pancrazzi\inst{\ref{inaf-oato}} \and
C.~Sasso\inst{\ref{inaf-oac}} \and
D.~Spadaro\inst{\ref{inaf-oact}} \and
R.~Susino\inst{\ref{inaf-oato}} \and
D.~Telloni\inst{\ref{inaf-oato}} \and
L.~Teriaca\inst{\ref{mps}} \and
M.~Uslenghi\inst{\ref{inaf-iafs}} \and
Y.-M.~Wang\inst{\ref{nrl}} \and
A.~Bemporad\inst{\ref{inaf-oato}} \and
G.~Capobianco\inst{\ref{inaf-oato}} \and
M.~Casti\inst{\ref{catholic},\ref{ass}} \and
M.~Fabi\inst{\ref{uniurb}} \and
F.~Frassati\inst{\ref{inaf-oato}} \and
F.~Frassetto\inst{\ref{ifn-cnr},\ref{ass}} \and
S.~Giordano\inst{\ref{inaf-oato}} \and
C.~Grimani\inst{\ref{uniurb}} \and
G.~Jerse\inst{\ref{inaf-oats}} \and
E.~Magli\inst{\ref{polito}} \and
G.~Massone\inst{\ref{inaf-oato}} \and
M.~Messerotti\inst{\ref{inaf-oats}} \and
D.~Moses\inst{\ref{nasa}} \and
M.-G.~Pelizzo\inst{\ref{ieiit-cnr}} \and
P.~Romano\inst{\ref{inaf-oact}} \and
U.~Sch\"uhle\inst{\ref{mps}} \and
A.~Slemer\inst{\ref{ifn-cnr}} \and
M.~Stangalini\inst{\ref{asi},\ref{ass}} \and
T.~Straus\inst{\ref{inaf-oac}} \and
C.~A.~Volpicelli\inst{\ref{inaf-oato}} \and
L.~Zangrilli\inst{\ref{inaf-oato}} \and
P.~Zuppella\inst{\ref{ifn-cnr},\ref{ass}} \and
L.~Abbo\inst{\ref{ass}} \and
F.~Auch\`ere\inst{\ref{ias}} \and
R.~Aznar Cuadrado\inst{\ref{mps}} \and
A.~Berlicki\inst{\ref{uniwrocl},\ref{asu}} \and
R.~Bruno\inst{\ref{inaf-iaps}} \and
A.~Ciaravella\inst{\ref{inaf-oapa}} \and
R.~D'Amicis\inst{\ref{inaf-iaps}} \and
P.~Lamy\inst{\ref{lab_atm}} \and
A.~Lanzafame\inst{\ref{unict},\ref{ass}} \and
A.~M.~Malvezzi\inst{\ref{unipv}} \and
P.~Nicolosi\inst{\ref{unipd}} \and
G.~Nistic\`o\inst{\ref{unical}} \and
H.~Peter\inst{\ref{mps}} \and
C.~Plainaki\inst{\ref{asi},\ref{ass}} \and
L.~Poletto\inst{\ref{ifn-cnr}} \and
F.~Reale\inst{\ref{unipa},\ref{ass}} \and
S.~K.~Solanki\inst{\ref{mps}} \and
L.~Strachan\inst{\ref{nrl}} \and
G.~Tondello\inst{\ref{unipd}} \and
K.~Tsinganos\inst{\ref{uniat}} \and
M.~Velli\inst{\ref{unica}} \and
R.~Ventura\inst{\ref{inaf-oact}} \and
J.-C.~Vial\inst{\ref{ias}} \and
J.~Woch\inst{\ref{mps}} \and
G.~Zimbardo\inst{\ref{unical},\ref{ass}}}

\institute{University of Firenze, Italy\label{unifi} -- \email{marco.romoli@unifi.it} \and
INAF -- Associate Scientist\label{ass} \and
INAF -- Astronomical Observatory of Capodimonte, Naples, Italy\label{inaf-oac} \and
University of Catania, Italy\label{unict} \and
INAF -- Astrophysical Observatory of Catania, Italy\label{inaf-oact} \and
IFN-CNR, Padova, Italy\label{ifn-cnr} \and
Max-Planck-Institut f\"ur Sonnensystemforschung, G\"ottingen, Germany\label{mps} \and
Predictive Science Inc., San Diego, California, US\label{psi} \and
INAF -- Astrophysical Observatory of Torino, Italy\label{inaf-oato} \and
Astronomical Institute of the Czech Academy of Sciences, Ond\v{r}ejov, Czech Republic\label{asu} \and
University of Torino, Italy\label{unito} \and
University of Padova, Italy\label{unipd} \and
INAF -- Institute for Space Astrophysics and Cosmic Physics, Milan, Italy\label{inaf-iafs} \and
Naval Research Laboratory, Washington DC, US\label{nrl} \and
Catholic University of America, Washington DC, US\label{catholic} \and
University of Urbino, Italy\label{uniurb} \and
INAF -- Astrophysical Observatory of Trieste, Italy\label{inaf-oats} \and
Politecnico di Torino, Italy\label{polito} \and
NASA HQ, Washington DC, US\label{nasa} \and
IEIIT-CNR, Padova, Italy\label{ieiit-cnr} \and
ASI -- Agenzia Spaziale Italiana, Roma, Italy\label{asi} \and
Institut d'Astrophysique Spatiale, Paris, France\label{ias} \and
University of Wroc\l{}aw - Astronomical Institute, Poland\label{uniwrocl} \and
INAF -- Institute for Space Astrophysics and Planetology, Rome, Italy\label{inaf-iaps} \and
INAF – Astronomical Observatory of Palermo, Italy\label{inaf-oapa} \and
Laboratoire Atmosphères, Milieux et Observations Spatiales, Guyancourt, France\label{lab_atm} \and
University of Pavia, Italy\label{unipv} \and
University of Calabria, Italy\label{unical} \and
University of Palermo, Italy\label{unipa} \and
University of Athens, Greece\label{uniat} \and
University of California, Los Angeles, California, US\label{unica}}

\titlerunning{First light observations with Metis - Solar Orbiter coronagraph}
\authorrunning{Romoli et al.}

\date{}

\abstract{The investigation of the wind in the solar corona initiated with the observations of the resonantly scattered ultraviolet emission of the coronal plasma obtained with UVCS-SOHO, that was designed to measure the wind outflow speed by applying the Doppler dimming diagnostics.
Metis on Solar Orbiter complements the UVCS spectroscopic observations, performed during solar activity cycle 23, by simultaneously imaging the polarized visible light and the \ion{H}{i}~Lyman-$\alpha$ corona in order to obtain high-spatial and temporal resolution maps of the outward velocity of the continuously expanding solar atmosphere.
The Metis observations, on May~15, 2020, provide the first \ion{H}{i}~Lyman-$\alpha$ images of the extended corona and the first instantaneous map of the speed of the coronal plasma outflows during the minimum of solar activity and allow us to identify the layer where the slow wind flow is observed.
The polarized visible light (580-640~nm), and the ultraviolet \ion{H}{i}~Ly$\alpha$ (121.6~nm) coronal emissions, obtained with the two Metis channels, are combined in order to measure the dimming of the UV emission relative to a static corona.
This effect is caused by the outward motion of the coronal plasma along the direction of incidence of the chromospheric photons on the coronal neutral hydrogen.
The plasma outflow velocity is then derived as a function of the measured Doppler dimming.
The static corona UV emission is simulated on the basis of the plasma electron density inferred from the polarized visible light.
This study leads to the identification, in the velocity maps of the solar corona, of the high-density layer about $\pm10\degr$ wide, centered on the extension of a quiet equatorial streamer present at the East limb -- the coronal origin of the heliospheric current sheet -- where the slowest wind flows at about $160\pm18$~km~s$^{-1}$ from 4~$R_\odot$ to 6~$R_\odot$.
Beyond the boundaries of the high-density layer, the wind velocity rapidly increases, marking the transition between slow and fast wind in the corona.}


\maketitle

\section{Introduction\label{sect:intro}}

The first direct observations of the solar wind at coronal heights were obtained with the Ultraviolet Coronagraph Spectrometer \citep[UVCS;][]{Kohl1995, Kohl1997} on the Solar and Heliospheric Observatory \citep[SOHO;][]{domingo1995}. Coronal outflows were detected out to about 5~$R_\odot$ by adopting spectroscopic techniques based on the Doppler dimming of the resonantly scattered ultraviolet light emitted by coronal ions and atoms.
The main UVCS results on the solar wind, concerning the measurement of the outflow speed in coronal hole and streamer regions and the observational evidence of energy deposition in the coronal hole regions, are reported in several reviews \cite[e.g.,][]{antonucci2012,abbo2016,cranmer2017,antonucci2020a}.

UVCS allowed the detection of the \ion{H}{i}~Ly$\alpha$ light at very high spatial and spectral resolution but in a narrow instantaneous field of view.
Therefore, maps of the hydrogen atoms outflow velocity in the corona could only be reconstructed, under the hypothesis of steady conditions, from data collected at different times and without a continuous spatial coverage of a given annulus of the corona \citep[e.g.,][]{dolei2018}.
The expansion speed of the coronal plasma was also derived from the dimming observed with UVCS in the ultraviolet \ion{O}{vi} doublet (103.2, 103.7~nm), emitted by five-time ionized oxygen ions, O$^{5+}$.
The expansion rate of the minor component of the solar wind formed by oxygen ions was extensively studied, out to 5~$R_\odot$, along the streamer boundaries \citep[e.g.,][]{frazin2003,susino2008} and in polar coronal holes where the energy deposition process responsible for the acceleration of the fast wind is more evident \citep[]{telloni2007a}.

Metis, the coronagraph of the Solar Orbiter mission \citep[][]{muller2020} was designed with the purpose of widening the knowledge previously acquired with UVCS.
This is achieved by privileging imaging at high temporal and spatial resolution of the full off-limb corona in the polarized visible light (VL) in the band 580-640~nm, and in the ultraviolet \ion{H}{i}~Lyman-$\alpha$ line (121.6~nm), in order to trace the global dynamics and evolution of the proton component in the solar atmosphere from 1.7~$R_\odot$ to 9~$R_\odot$ \citep[][]{antonucci2020b}.

\begin{figure*}
	\centering
	\includegraphics[width=\textwidth]{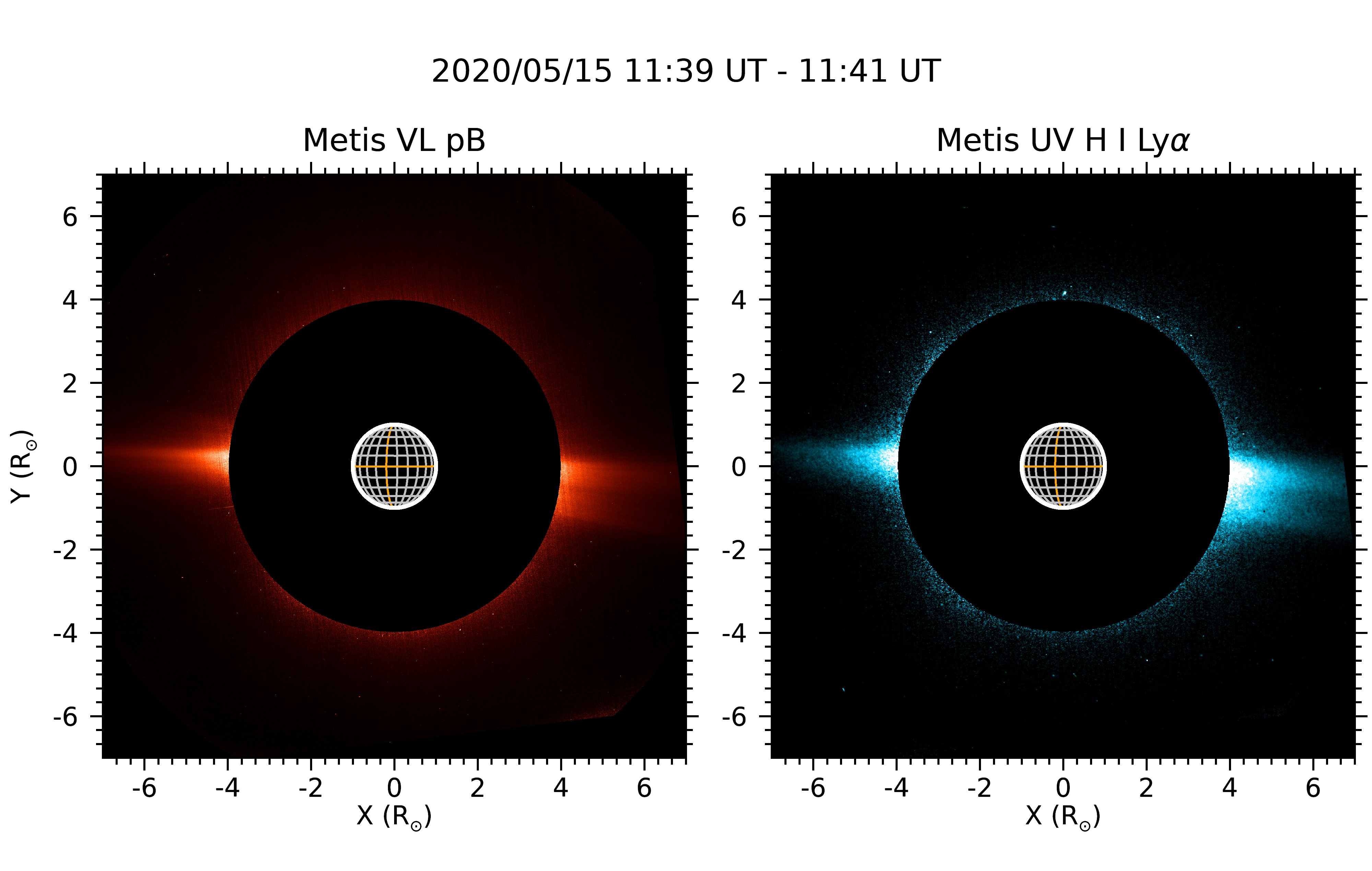}
	\caption{\label{fig:1}Visible-light polarized brightness within 580-640~nm band (left panel) and ultraviolet \ion{H}{i}~Ly$\alpha$ emission at 121.6~nm (right panel) of the solar corona observed from a vantage point of 0.64~AU on May~15, 2020, at 11:40~UT, within a field-of-view ranging from 3.8~$R_\odot$ to 7~$R_\odot$.}
\end{figure*}

The coronal \ion{H}{i}~Lyman-$\alpha$ emission, primarily due to resonance scattering of the chromospheric \ion{H}{i}~Lyman-$\alpha$ photons by part of the residual neutral hydrogen present in the solar corona \citep[][]{gabriel1971}, depends on the outward velocity of the outflowing plasma relative to the source of the photons at the base of the corona \citep[][]{hyder1970,beckers1974,withbroe1982,noci1987}.
Hence, the Doppler dimming of the UV line induced by this relative motion allows us to map, at high spatial resolution and temporal cadence, the speed of the solar wind in the corona over an extended field of view.
In order to measure this effect, the observed UV emission is compared with that for a static corona expected on the basis of the polarized VL emission, detected simultaneously to the UV emission, which provides a measure of the electron density of the plasma.
Hence, Metis provides instantaneous high temporal and spatial resolution maps of the solar atmosphere which is continuously expanding in the form of the solar wind.
The aim here is to study the wind in its initial phase of outward propagation in the inner corona when it flows along the open field lines, strongly modulated by the topology of the coronal magnetic field, and at radial distances beyond 4-5~$R_\odot$, where the plasma is not confined any longer thus the solar atmosphere is fully permeated by wind outflows.

Due to the high rate of charge exchange out to about 10~$R_\odot$ between hydrogen atoms and protons in the coronal plasma \citep[e.g.,][]{olsen1994,allen1998}, the neutral hydrogen outflow is representative of that of protons, which are the major component of the solar wind.
However, in faint regions such as polar coronal holes, beyond a certain height estimated to be approximately 4~$R_\odot$, the \ion{H}{i}~Ly$\alpha$ emission can become less intense than the interplanetary \ion{H}{i}~Ly$\alpha$ emission \citep[e.g.,][]{nakagawa2008,spadaro2017}.
Thus according to these studies, in polar holes there is a limit in height to the detection of  the solar wind via Doppler dimming of the \ion{H}{i}~Ly$\alpha$ line.

Along the radial extension of coronal streamers, where the current sheet separates open magnetic field lines of opposite polarities, the expansion rate of the outer corona can also be traced by streamer blobs, tiny flux ropes detached via reconnection events, due either to reconnection among streamer loops or to interchange reconnection between closed loops and open flux at the coronal hole-streamer boundaries \citep[][]{sheeley2009,wood2020}.
Blobs are travelling outward approximately at the same speed derived on the basis of Doppler dimming for the coronal wind plasma \citep[e.g.,][]{abbo2010}.

Results on the dynamics and structure of the solar wind in the corona are summarized in recent reviews \citep[e.g.,][]{abbo2016,cranmer2017,antonucci2020a}.

\section{Metis coronal observations on May 15, 2020}

The Metis coronagraph obtained its first images of the solar corona during the Solar Orbiter commissioning phase on May~15, 2020, from a 0.64~AU vantage point \citep[the Metis commissioning activities are reported in][]{romoli2021}.
The Solar Orbiter-Sun direction formed an angle of $11.4\degr$ West relative to the Earth-Sun direction and the spacecraft was at $4.3\degr$ North relative to the solar equatorial plane.
A coronal annular region from 3.8 to 7~$R_\odot$ was observed at the limb in the time interval 11:39:25-11:41:00~UT.
During this observation, the solar corona was imaged for the first time simultaneously both in polarized visible light (VL), within a band ranging from 580~nm to 640~nm, emitted by free electrons scattering of the photospheric light, and in the $121.6\pm10$~nm \ion{H}{i}~Ly$\alpha$ ultraviolet (UV) light, which is caused mainly by neutral hydrogen resonant scattering of chromospheric \ion{H}{i}~Ly$\alpha$ photons (Figure~\ref{fig:1}).
The acquisition of polarized visible light ensures that the signal is due to electron scattering (K-corona) and not to dust particles (F-corona), whose effect becomes more and more important with helio-distance.

Starting at 11:20~UT on May~15, 2020, a series of two acquisitions with the Metis coronagraph was commanded: two polarized-brightness sequences of four polarimetric frames with detector integration time equal to 15~s and 30~s, respectively, for the VL channel, and two sets of 6~UV images, averaged over 8 and 6 frames, each with integration time of 8~s and 16~s, respectively, for the UV channel.
Both VL and UV detectors were configured with a $1\times1$ pixel binning, corresponding to an image scale on the plane of the sky of about 4300~km and 8600~km, respectively.
Metis full width half maximum (FWHM) spatial resolution spans throughout the field-of-view from 1.5 to 3~pixels in the VL channel and from 4 to 6.5~pixels in the UV channel \citep[][]{dadeppo2021}.

Figure~\ref{fig:1} (left) depicts, the visible-light polarized brightness obtained by combining the polarimetric images with detector integration time of 30~s and by using the theoretical M\"uller demodulation matrix \citep[e.g.,][]{fineschi2005,casti2019}.
The corresponding UV~\ion{H}{i}~Ly$\alpha$ corona imaged with detector integration time of 16~s is shown on the right.
At a helio-distance equal to 0.64~AU, the observed field of view ranged from 3.8 to 7~$R_\odot$.

The set of images was processed and calibrated on ground as follows: bias and dark subtraction, flat-field and vignetting correction, exposure-time normalization, and radiometric calibration.
The bias and dark images for both detectors were acquired in flight, whereas the flat-field and vignetting images as well as the radiometric parameters, used to convert the image values from digital number to the physical units, were measured on ground during the laboratory calibrations \citep[][]{antonucci2020b}.
For the vignetting correction of the UV images, an improved vignetting function was obtained by combining the laboratory function with images acquired on-board before the opening of the Metis external door.
The in-flight radiometric calibration is currently in progress, based on a set of standard VL and UV stars that can cover the various portions of the Metis field-of-view, so giving information on the radiometric response of the instrument all over its field-of-view.
For this reason, the wind velocities are analyzed in detail in the region where the UV intensity of the alpha-Leo star was recorded during its transit across the Metis field-of-view, thus giving the possibility to refine the correction of the UV intensity in the equatorial region observed at the East limb.
In that region, the observed non-uniformity ($\sim20\%$) of the UV channel leads to an uncertainty in the outflow velocity measurement of $\pm4$~km~s$^{-1}$.
The set of four polarimetric VL images were eventually demodulated with the M\"uller matrix derived from on-ground calibrations to derive the polarized-brightness image.

The ultraviolet image of the emission from neutral hydrogen atoms in Figure~\ref{fig:1} (right panel) is the first \ion{H}{i}~Ly$\alpha$ image of the outer solar corona ever obtained.
The brightness distribution in the field-of-view reproduces very well that of the polarized visible light (left panel) with the two bright equatorial streamers and fainter polar regions.
The streamers appear significantly bright all the way to the upper limit of the field-of-view, showing the presence of emitting neutral hydrogen at large distances from the Sun.

The visible light polarized brightness image clearly shows the characteristic configuration of the solar corona during the minimum phase of magnetic activity, when the global scale magnetic field is predominantly dipolar and confines the plasma mostly near the equator, giving rise to a quasi-equatorial streamer belt.
The polar regions, where the fast wind is flowing along the open magnetic field lines, exhibit a fainter brightness.
The classic solar minimum configuration of the corona observed with Metis in May closely resembles the persistent dipolar configuration observed with SOHO during 1996, that is, one full magnetic cycle earlier when the streamer belt was only slightly warped around the equatorial plane.
At the limb and on the plane of the sky, this is viewed as two almost equatorial streamers stretching out to form the coronal/heliospheric current sheet surrounded by a plasma enhancement (Figure~\ref{fig:1}).
The 1996 solar minimum corona, studied extensively in the first year of the SOHO coronagraphs observations, is described, for example, by \citet[][]{schwenn1997} who combined the observations of the three LASCO coronagraphs \citep[][]{brueckner1995}, thus ensuring an unprecedented continuous coverage of the off-limb corona from 1.1 to 32~$R_\odot$.

The similarity with the 1996 minimum is also confirmed by the synoptic chart of the coronal magnetic field computed on the basis of the Wilcox Stanford Observatory, relative to Carrington Rotation (CR) 2230 (\url{http://wso.stanford.edu}).
The map shows that, at the source surface at 3.25~$R_\odot$, the neutral line, which outlines with good approximation the source of the heliospheric current sheet, is laying close to the solar equator.

\begin{figure*}
	\centering
	\includegraphics[width=\textwidth]{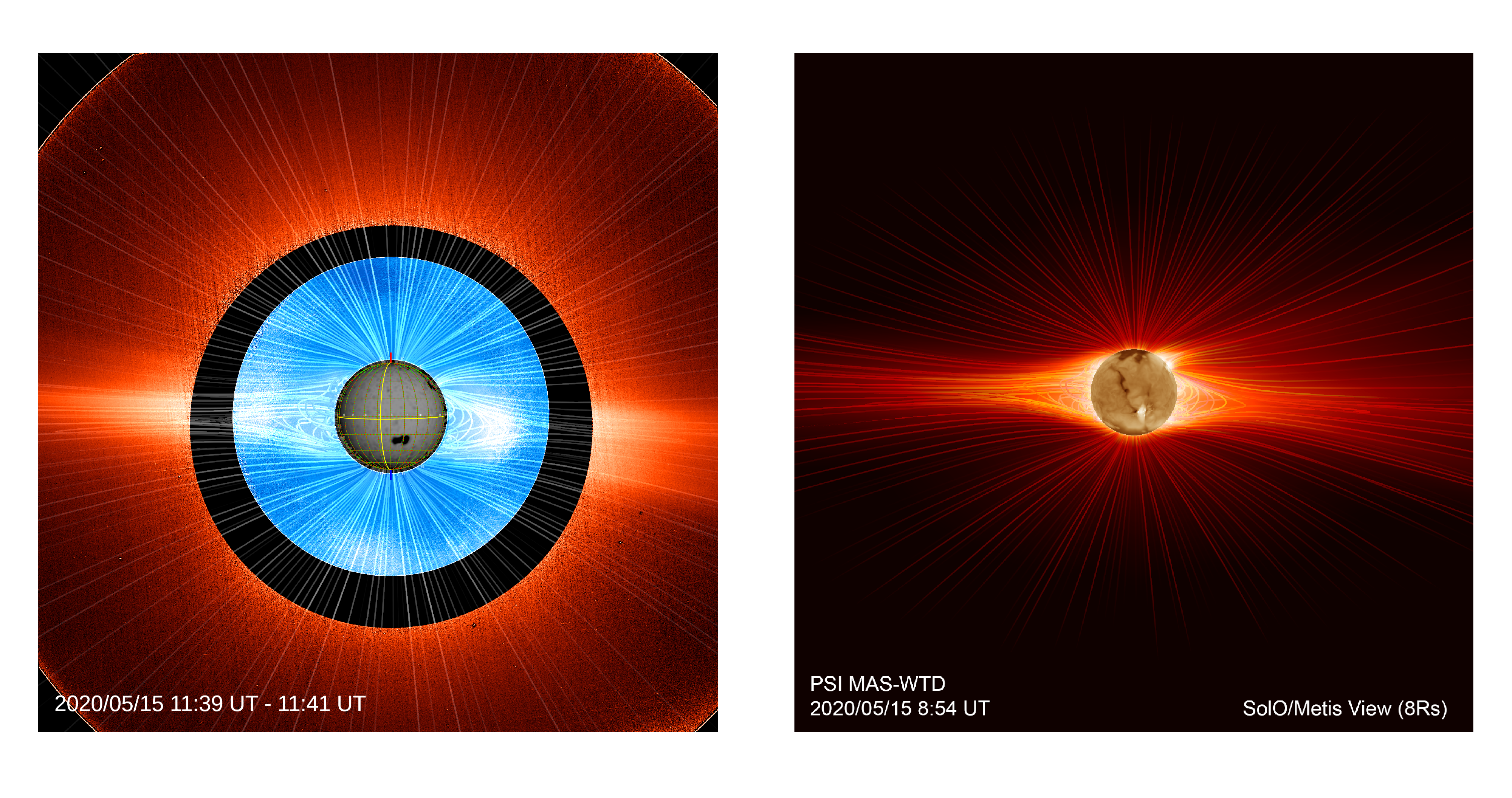}
	\caption{\label{fig:2}Composite image of the outer corona, observed with the Metis coronagraph in polarized visible light, and of the medium and inner corona observed at the Mauna Loa Solar Observatory, with the inner magnetic boundary condition ($B_r$) of the PSI-MHD model at 1~$R_\odot$ (left panel).
	Field lines traced near the plane-of-sky from the PSI-MHD simulation are overlaid over the composite image.
The right panel shows the same field lines, but now with forward modeled SDO/AIA 193~\AA\ and polarized brightness observables computed directly from the simulation.}
\end{figure*}

\begin{figure}
	\centering
	\includegraphics[width=0.75\columnwidth]{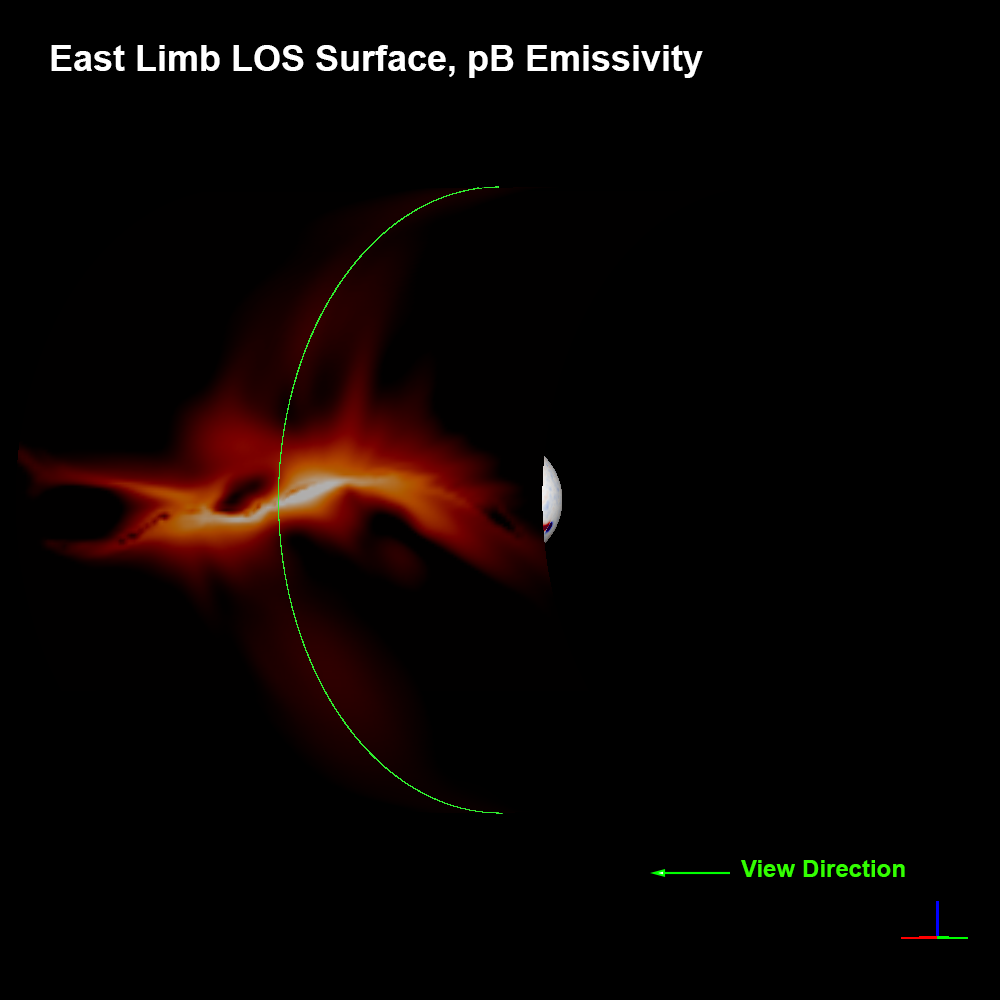}
	\caption{\label{fig:3}Line-of-sight (LOS) surface of the pB emissivity in the extended corona, calculated at 5~$R_\odot$ at the East limb on the basis of the Predictive Science MHD model, highlighting the slightly warped equatorial density sheet surrounding the heliospheric current sheet.}
\end{figure}

For context, we also use a 3D magnetohydrodynamic (MHD) model of the solar corona, generated by Predictive Science Incorporated (PSI).
This particular calculation employs the same MHD model and Wave-Turbulence-Driven (WTD) heating approach described by \citet[][]{mikic2018}, although at slightly lower resolution.
To best represent the first month or so of initial measurements in one run, the magnetic boundary conditions for the PSI-MHD model are based on SDO/HMI \citep[][]{scherrer2012} synoptic map data, with magnetic measurements spanning roughly the period from May~31 to June~26 (mostly CR2231 and some CR2232).
Figure~\ref{fig:2} shows diagnostics from the model, including field lines traced roughly from the plane-of-sky overlaid over Metis and Mauna Loa polarized brightness observations (left panel), and the same field lines overlaid over forward modeled SDO/AIA \citep[][]{lemen2012} and polarized brightness observables out to $\pm8~R_\odot$ (right panel).
The field lines indicate the coronal regions of closed field lines, which confine the plasma in the inner part of the streamer equatorial belt, and the fully open field regime in the Metis field of view.
That is, at the heights observed with Metis, the field does not confine the coronal plasma anymore and is instead dragged outward by the solar wind.
In Figure~\ref{fig:3}, the surface computed at 5~$R_\odot$ represents the line-of-sight path along which the polarized brightness emissivity of the corona is observed, computed on the basis of the MHD model (the emissivity is depicted on a cylinder in/out of the plane of the sky rotated by $\pm45\degr$).
The modelled emissivity confirms that the equatorial density sheet is quite close to the equator.

\section{Data analysis and outflow velocity diagnostic technique\label{sect:data_analysis}}

The simultaneous detection of the solar corona emission in polarized visible light and the \ion{H}{i}~Ly$\alpha$ line allows us to use the Doppler dimming technique \citep[][]{withbroe1982,noci1987} to derive the radial outflow velocities of neutral hydrogen atoms in the solar corona.
As anticipated in Sect.~\ref{sect:intro}, this diagnostic technique is essentially based on the comparison of the \ion{H}{i}~Ly$\alpha$ coronal emission, synthesized on the basis of the simultaneous electron density measurements derived from the polarized brightness data, with the UV emission observed with Metis, as described in \citet[][]{antonucci2020b}.
The capability of mapping the coronal outflow velocity using combined VL and UV observations was recently illustrated by \citet[][]{dolei2018,dolei2019}, and \citet[][]{capuano2021}.
We refer the reader to these authors for a detailed description of the derivation of solar wind outflow velocity maps.

For the computation of the synthetic coronal UV emission, information is required about physical quantities on which the Ly$\alpha$ intensity depends: electron density $n_{e}$, electron temperature $T_{e}$ and kinetic temperature of the hydrogen atoms, $T_{\ion{H}{i}}$, equivalent to the atom velocity distribution of the coronal plasma along a given direction. The values adopted for such plasma parameters are discussed below.

\paragraph {Electron density} The 2D distribution of the coronal electron density $n_{e}$ is derived from Metis polarized brightness map by means of the inversion technique developed initially by \citet[][]{vandehulst1950} and by adopting the expressions described and used in \citet[][]{dolei2015,dolei2018}.
This quantity is derived with an uncertainty $\leq10\%$, leading to an uncertainty $\leq10$~km~s$^{-1}$ in the outflow velocity of the solar wind.

\paragraph {Electron temperature} The radial variation of $T_{e}$ in the outer coronal region is assumed to be an average of the $T_{e}$ radial profiles derived at solar minimum by \citet[][]{gibson1999} for equatorial regions and by \citet[][]{vasquez2003} for polar regions, according to the same approach used by \citet[][]{dolei2018}.

\paragraph {Hydrogen kinetic temperature} The $T_{\ion{H}{i}}$ kinetic temperatures relative to a solar minimum epoch are obtained from the database created by considering a large collection of kinetic temperatures, derived from UVCS Ly$\alpha$ spectroscopic observations over different temporal windows \citep[][]{dolei2016,dolei2019}.
It is worth mentioning that the kinetic temperature is deduced from the $\ion{H}{i}$~Ly$\alpha$ line broadening observed with UVCS, determined by the velocity distribution of the hydrogen atoms along the line of sight, which includes contributions due to non-thermal mechanisms, for instance, due to energy deposition across the coronal magnetic field.
The $\ion{H}{i}$ kinetic temperature distribution in the corona can be anisotropic, as found in coronal holes on the basis of the UVCS data analysis \citep[see, e.g.,][]{kohl1998,cranmer1999,antonucci1999,antonucci2000}.
In the polar regions, for what concerns the O$^{5+}$ ions,  the lowest degree of the kinetic temperature anisotropy still compatible with the UVCS observations reaches a maximum value approximately equal to 15, at 2.9~$R_\odot$.
The anisotropy ratio is then relaxing toward $\sim2$ at larger heliocentric distances, beyond about 3.7~$R_\odot$ \citep[][]{telloni2007b}.
By employing a model of the magnetic field representing the solar minimum corona and the wind speed regulated by mass-flux conservation, in a series of papers Raouafi and Solanki argue that the kinetic temperature of the O$^{5+}$ ions  across the magnetic field can be reproduced with good accuracy by taking into account the electron density along the line of sight as deduced from SOHO data and without necessarily invoking a degree of anisotropy \citep[][]{raouafi2003,raouafi2004,raouafi2006}.
These authors also suggest that isothermal conditions for heavy ions, protons and electrons cannot be ruled out.

In our study, given the $T_{\ion{H}{i},\perp}$  (provided by the UVCS database), both limits of isotropic ($T_{\ion{H}{i},\parallel}=T_{\ion{H}{i},\perp}$) and fully an\-i\-so\-trop\-ic ($T_{\ion{H}{i},\parallel}=T_e$) distribution of $T_{\ion{H}{i}}$ are considered.
That is, the kinetic temperature of the hydrogen atoms across the radial direction ($T_{\ion{H}{i},\perp}$) is equal to that determined by the UVCS observations during solar minimum and its value is the maximum one, whilst along the radial direction ($T_{\ion{H}{i},\parallel}$) this is assumed to vary from the maximum value in the isotropic case, to a value of the hydrogen kinetic temperature equal to the electron temperature ($T_e$) in the case of the maximum anisotropy degree.

\paragraph {Chromospheric radiation} As for the chromospheric $\ion{H}{i}$~Ly$\alpha$ radiation exciting the hydrogen atoms in the corona, we assume uniform disc brightness, with a line intensity equal to $8.16\times 10^4$~erg~cm$^{-2}$~s$^{-1}$~sr$^{-1}$ ($5.0\times 10^{15}$~ph~cm$^{-2}$~s$^{-1}$~sr$^{-1}$), derived from SOHO/SUMER observations by \citet[][]{lemaire2015}, and adopt the analytical line profile proposed by \citet[][]{auchere2005} \citep[see also][]{dolei2018,dolei2019}.
\citet[][]{dolei2019} also investigated the effect of a non-uniform chromospheric Ly$\alpha$ radiation on determining the coronal hydrogen atoms outflow velocity.
They found that the uniform-disc approximation leads to overestimated velocities in the polar and mid-latitude coronal regions, up to some tens km~s$^{-1}$ closer to the Sun, and to slightly underestimated velocities at equatorial latitudes.
These differences decrease at higher heliocentric distances and the non-uniform radiation condition progressively approaches the uniform-disc approximation.

For the computation of the synthetic coronal Ly$\alpha$ line intensity, we calculated the emissivity of the line for each volume element along the line of sight.
Then, we obtained the expected intensities by adding up the emissivities of all these elements along a total length of 10~$R_\odot$ ($\pm5~R_\odot$ with respect to the plane of the sky), sampled in steps of 0.5~$R_\odot$.
This length along the line of sight is consistent with the value adopted to derive the electron densities by the inversion of the polarized brightness data.
Before applying the Doppler dimming diagnostics, in order to improve the UV image statistics and increase the signal-to-noise ratio, the calibrated images were further reduced with a $4\times 4$ pixel binning.
The resulting signal-to-noise ratio is greater than 5 in a region approximately $\pm20\degr$ wide across the equatorial belt, thus including the streamers observed at the East and West limb of the Sun.
Due to the performed data binning the polar angle element is of 1 degree on average.

\section{Outflow velocity results in the outer corona}

\begin{figure*}
	\centering
	\includegraphics[width=0.475\textwidth]{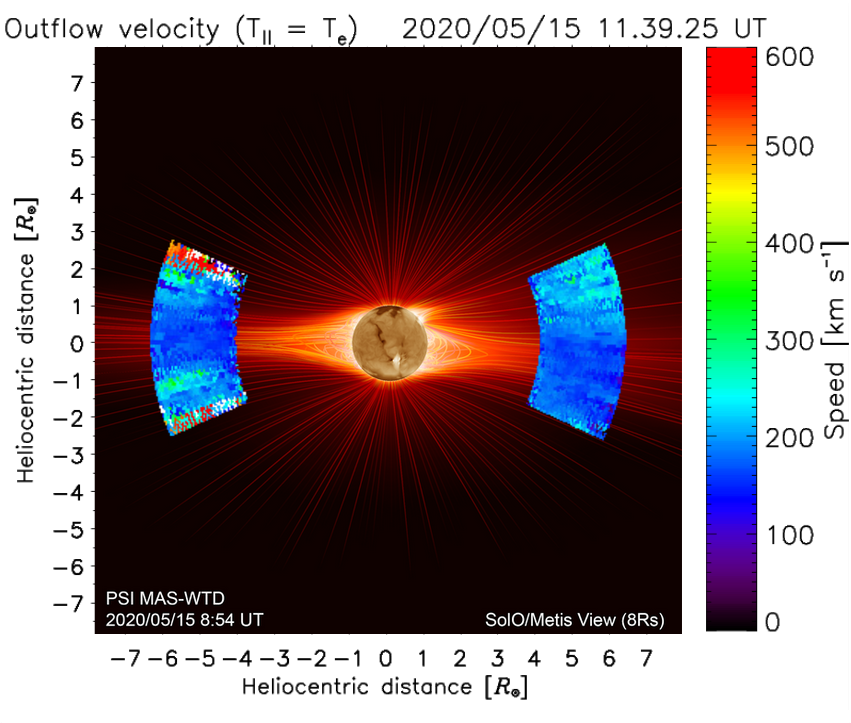}
	\includegraphics[width=0.475\textwidth]{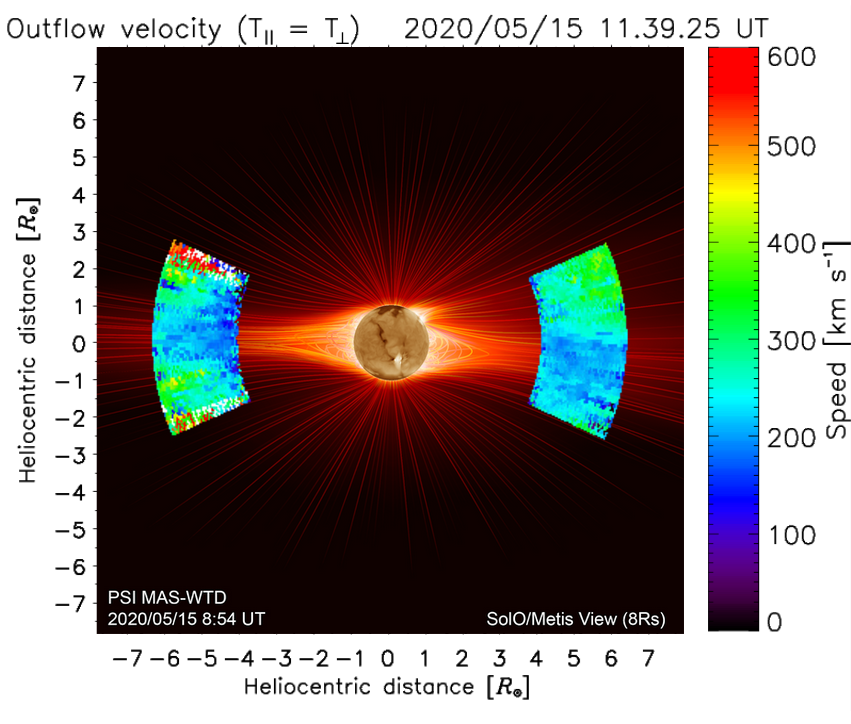}
	\caption{\label{fig:4}\ion{H}{i} outflow velocity map, within the region where the signal-to-noise ratio is equal to 5, in the case of anisotropy, $T_{\ion{H}{i},\parallel}=T_e$ (left panel) and of isotropy $T_{\ion{H}{i},\parallel}=T_{\ion{H}{i},\perp}$ (right panel) of the neutral hydrogen kinetic temperature along the radial direction.}
\end{figure*}

In the first measurement of the outflow velocity of the plasma in the coronal region $\pm20\degr$ wide across the equator, obtained with Metis data (Figure~\ref{fig:4}), this quantity is computed by means of the code described in Sect.~\ref{sect:data_analysis}.
The velocity results are displayed within the latitude range limited to the latitudes where the signal-to noise ratio in the \ion{H}{i}~Ly$\alpha$ intensity is $\geq5$.
The map of the wind velocity is derived for values of the kinetic temperature of the hydrogen atoms in both anisotropic (left panel) and isotropic (right panel) cases.
At the equator the velocities computed for the isotropic case are higher by about 40~km~s$^{-1}$ than those derived if the coronal profiles of the scattering atoms are different in the direction parallel and perpendicular to the magnetic field lines.
Departing from the equatorial zone the velocity difference for the two cases tends to increase.
The map extends out to $6~R_\odot$, with the Doppler dimming computed along the line of sight from $-5~R_\odot$ to $+5~R_\odot$.
The velocity is measured with angular resolution of 1 degree.

\begin{figure*}
	\centering
	\includegraphics[width=\textwidth]{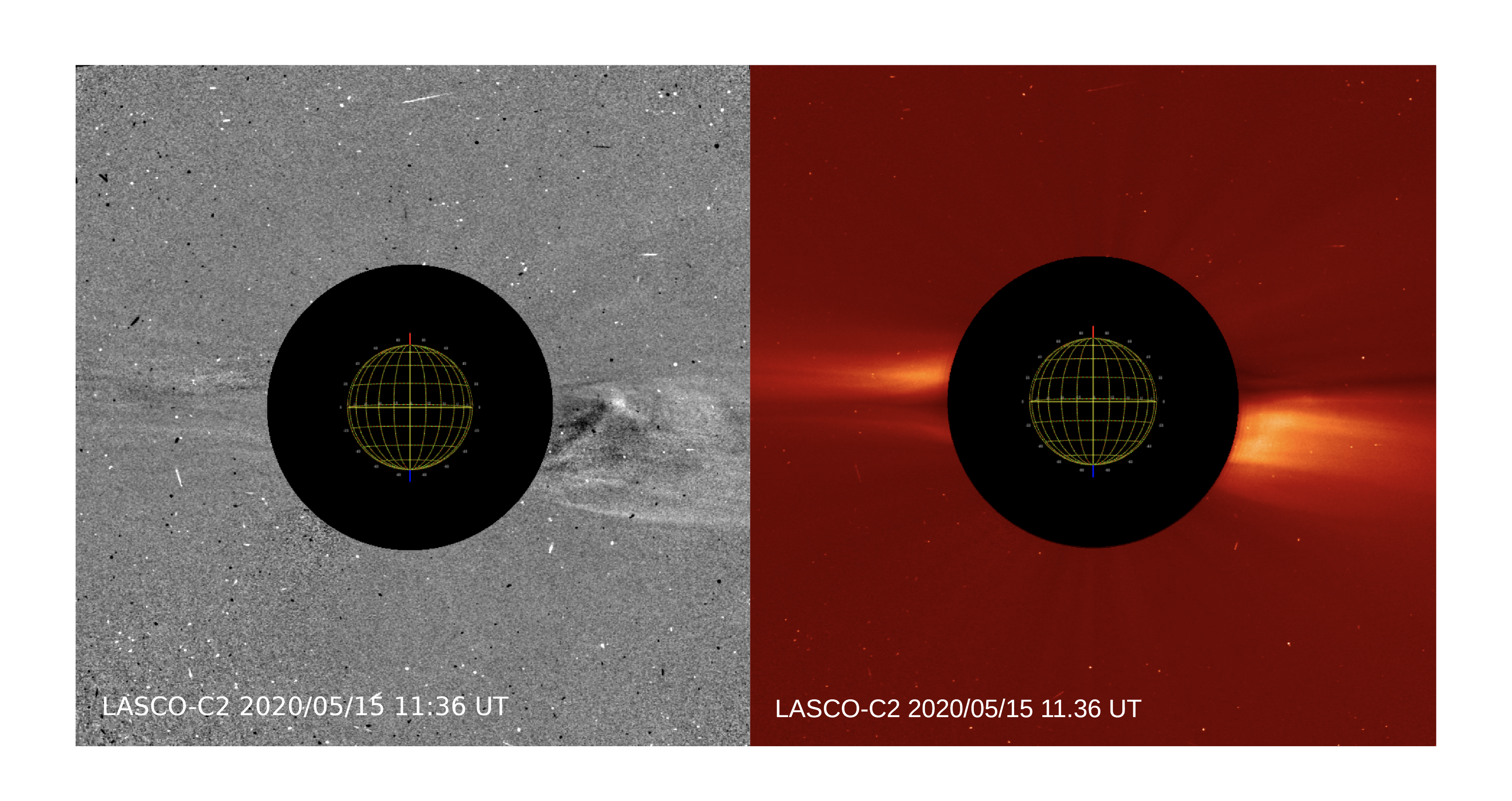}
	\caption{\label{fig:5}Coronal blob observed with LASCO C2 on SOHO, on May~15, 2020, at 11:36~UT, outlined in the image difference (left panel), entering the Metis field of view and moving along the West streamer.
	The right panel shows the corona observed with LASCO C2.}
\end{figure*}

At the West limb, the LASCO C2 observations show that a tiny streamer blob is entering the field of view of Metis as shown in the image observed at 11:36 UT of Figure~\ref{fig:5}, and is moving along the West limb streamer, which is characterized by a wider latitudinal structure than the streamer at the East limb.
This might be the result either of the transient phenomena occurring at the time of the Metis observation or of a line-of-sight effect of a more warped streamer equatorial belt.
Since the Solar Orbiter-Sun direction forms an angle of $11.4\degr$ West relative to the LASCO (L1)-Sun direction, the Metis and LASCO planes of the sky are separated by the same angle, thus, although slightly offset, approximately the same coronal features are observed with the two coronagraphs at the West limb.

\begin{figure}
	\centering
	\includegraphics[width=0.5\columnwidth]{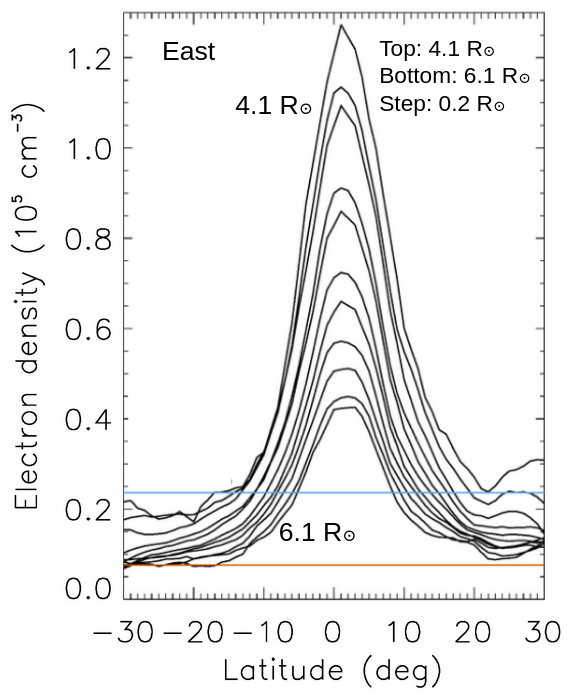}
	\caption{\label{fig:6}Latitudinal density profiles relative to the streamer observed at the East limb (polar angle 90$\degr$, equator).
	The horizontal lines represent typical coronal-hole densities \citep[e.g.,][]{spadaro2017} at 4.1~$R_\odot$ (upper line) and 6.1~$R_\odot$ (lower line).}
\end{figure}

The streamer observed at the East limb appears as a compact structure, well-defined in latitude, that outlines the projection on the plane of the sky of the surface dividing positive/negative polarity field lines of the northern/southern hemisphere.
This bright denser plasma layer, embedding the current sheet, is approximately delimited by the latitudes corresponding to the $1/e$ drop in density (Figure~\ref{fig:6}) and UV \ion{H}{i}~Ly$\alpha$ at about $\pm10\degr$ from the equator.
The width so defined is probably wider than that of the real plasma sheet due to the integration along the line of sight of the slightly tilted emissivity sheet shown in Figure~\ref{fig:3}.
\citet[][]{wang1998} report examples where the plasma sheet is seen almost edge on at the limb by LASCO C3 with a latitudinal width which can be as small as $\sim3\degr$, beyond 6~$R_\odot$.
Plasma sheets are likely to originate at the helmet streamer cusp where streamer loops may continually undergo interchange reconnection with the adjacent open flux, thus giving rise to these thin features \citep[][]{wang2012}.
The density sheet observed as extension of the West streamer is much wider, less dense and more extended toward the southern latitudes due to a more complex structure.
Thus, the East streamer is ideal to measure the plasma flows along and in proximity of the coronal-heliospheric current sheet, and for this reason is chosen in order to investigate the slowest component of the solar wind in the corona.

\begin{figure}
	\centering
	\includegraphics[width=0.75\columnwidth]{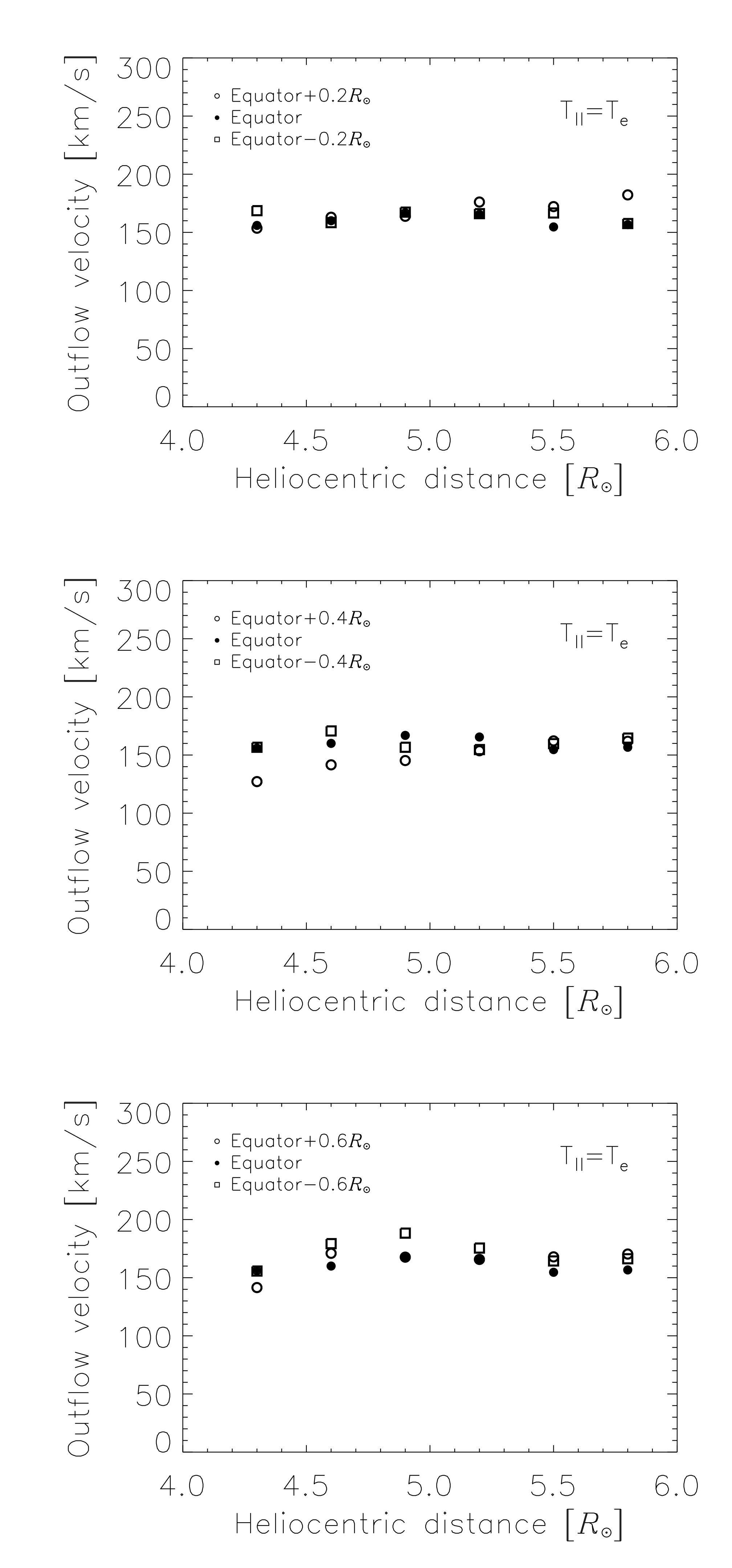}
	\caption{\label{fig:7}Outflow velocity along the axis of the East coronal streamer (full dots) at the equator; the velocity curves denoted by open circles and open squares are computed at a distance from the equator varying from $\pm0.2~R_\odot$ (upper panel) and $\pm0.4~R_\odot$ (mid panel) to $\pm0.6~R_\odot$ (lower panel), respectively at the North (+) and South (-).}
\end{figure}

The outflow velocity in the East streamer (Figure~\ref{fig:7}) is derived by applying the Doppler-dimming technique under the hypothesis that along the radial direction the \ion{H}{i} kinetic temperature is equal to the electron temperature, that is, in the anisotropic case.
If the outward velocity is derived in the equatorial region under the assumption of isotropy, its value increases of about 40~km~s$^{-1}$.
The assumed electron temperature is decreasing from 0.7~MK to 0.5~MK from 4~$R_\odot$ to 6~$R_\odot$ (see, Dolei, 2018).
Recently, in a balloon-borne investigation of a coronal streamer within 4 and 7~$R_\odot$, \citet{gopalswamy2021} found a higher electron temperature, $1.0\pm0.3$~MK, close to the value found for streamers by \citet{fineschi1998}, $1.1\pm0.3$~MK, but at a lower heliodistance, 2.7~$R_\odot$.
Across the radial direction, $T_{\ion{H}{i},\perp}$ is assumed to be equal to 1.6~MK, value determined by UVCS observations of the equatorial belt during solar minimum \citep[][]{dolei2016}.

The hypothesis of maximum anisotropy degree implies, as discussed earlier, that electron and hydrogen atoms/protons are in thermal equilibrium along the radial direction, that is, along the wind flow direction, whereas along the line-of-sight direction, approximately perpendicular to the coronal magnetic field, the \ion{H}{i} kinetic temperature is larger than the thermal temperature due to possible contribution of Alfv\'en waves and/or MHD turbulence fluctuations across the magnetic field.
The anisotropy degree of the hydrogen kinetic temperature cannot be accurately derived.
UVCS results point to kinetic temperatures perpendicular to the magnetic field larger than those in the parallel direction \citep[e.g.,][]{kohl1998,cranmer1999,li1999}.
Recently, \citet{cranmer2020}, on the base of a statistical analysis of coronal-hole data obtained with UVCS, concludes that almost-isotropic $T_{\ion{H}{i}}$ conditions are present in coronal holes, with an electron temperature of 1.2~MK, which is however quite higher than that directly measured at the base of coronal holes, 0.4~MK at 1.3~$R_\odot$ \citep[e.g.,][]{david1998,landi2008}.
These conclusions are in line with those reached by \citet{raouafi2003}, \citet{raouafi2004}, and \citet{raouafi2006}, reported in Sect.~\ref{sect:data_analysis}.

\begin{figure}
	\centering
	\includegraphics[width=0.75\columnwidth]{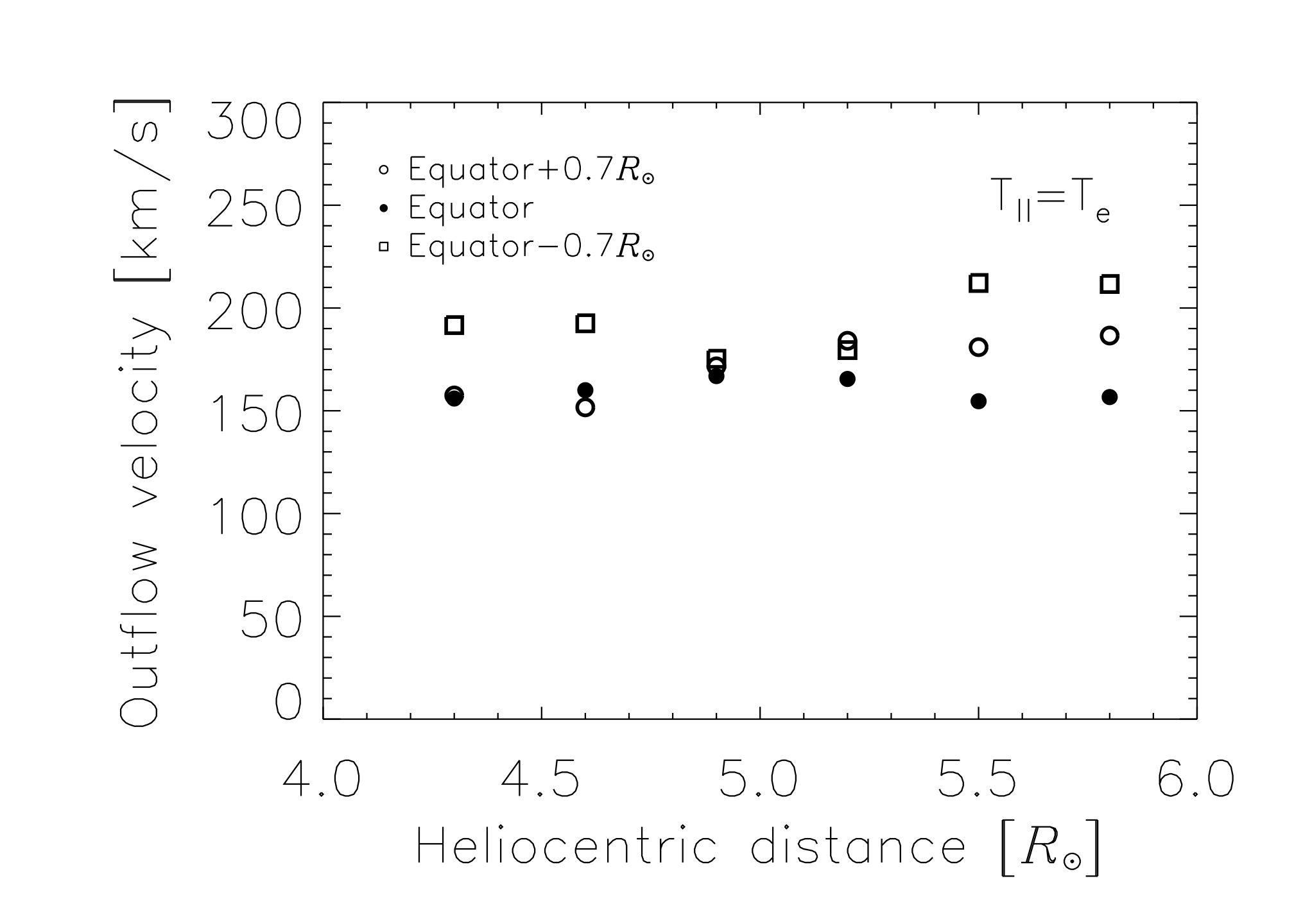}
	\caption{\label{fig:8}Outflow velocity along the axis of the East coronal streamer (full dots) at the equator, at $+0.7~R_\odot$ at North (open circles) and $-0.7~R_\odot$ at South (open squares).}
\end{figure}

In the equatorial sheet of enhanced density present at the East limb, the plasma is flowing at an almost uniform speed, $160\pm18$~km~s$^{-1}$, both in radial direction from 4.3~$R_\odot$ to 5.8~$R_\odot$ than in a zone $\pm0.6~R_\odot$ across the equator.
The velocities computed at a distance from the equator varying from $\pm0.2~R_\odot$ (upper panel) and $\pm0.4~R_\odot$ (mid panel) to $\pm0.6~R_\odot$ (lower panel) are plotted in Figure~\ref{fig:7}.

At $\pm0.7~R_\odot$, the distance from the streamer axis that corresponds to the $1/e$ density borders at 4~$R_\odot$, the outflow velocity is increasing from a value close to the equatorial one to about 185 and 210~km~s$^{-1}$ (northern and southern border, respectively) at 6~$R_\odot$ (see Figure~\ref{fig:8}). The outflow velocity is measured with angular resolution of 1 degree, and the values obtained with Doppler dimming have been averaged over 0.3~$R_\odot$ along the radial direction.
Figure~\ref{fig:7} shows that, within $\pm10\degr$ from the equator, a high-density solar wind is flowing quite uniformly at the lowest speed.
That is, in the observed annulus of the corona, the density layer embedding the current sheet is the site of a wind flowing at rather uniform speed along the radial direction, whereas beyond the high-density sheet, a sharp increase in wind velocity, e.g., of the order of tens~km~s$^{-1}$ deg$^{-1}$, is observed, which will be object of further study.

The fast wind in the polar regions, where the signal in the UV and VL channels is lower, will be studied when the in-flight radiometric calibrations will be complete.
Since the field of view of Metis extends to 9~$R_\odot$, it is of special interest to determine the height where in polar coronal holes the \ion{H}{i}~Ly$\alpha$ coronal emission becomes fainter  than the interplanetary \ion{H}{i}~Ly$\alpha$ one.
This level, which depends on the electron density and the level of dimming in the fast wind regime, at present is estimated to be approximately 4~$R_\odot$ (see Sect.~\ref{sect:intro}).
When polar coronal holes disappear with the increase of solar activity, the investigation of the outflows in polar regions will be probably possible over the entire Metis field of view.

\section{Discussion}
The present results on the wind speed, obtained thanks to the high-resolution observation of the full corona performed with Metis within an extended altitude range, show that the slow wind is flowing in the coronal plasma sheet, extension of an equatorial streamer, at an almost constant speed -- when its values are averaged over 0.3~$R_\odot$ in the flow direction -- within a layer $\pm~0.6~R_\odot$ wide across the equator and in the range from 4 to 6~$R_\odot$. In this coronal region the wind regime is supersonic.
On the basis of the electron temperature profile assumed in the analysis \citep[][]{gibson1999}, this condition is reached at about 4~$R_\odot$.
The transition to a faster wind occurs outside the plasma sheet of enhanced density and is characterized by a high velocity gradient in the $\approx~10\degr$ wide regions adjacent to the slow wind layer.

The UVCS spectrometer results, based on the Doppler dimming analyses of the \ion{O}{vi}~103.2~nm and \ion{O}{vi}~103.7~nm doublet and the \ion{H}{i}~Ly$\alpha$ line, although not covering continuously and simultaneously extended latitudinal regions of the solar minimum coronal streamers, have shown that the slowest wind is measured along the streamer axis where the O$^{5+}$ wind component is flowing at  $\approx100$~km~s$^{-1}$ \citep[e.g.,][]{habbal1997,strachan2002,frazin2003}.
During the declining phase of activity of cycle 23 along the extension of a mid-latitude streamer, \citet[][]{susino2008} find that at $5~R_\odot$ the hydrogen atoms are flowing outward at a much lower speed than the O$^{5+}$ ions  (60 and 135~km~s$^{-1}$, respectively).
However, in a solar minimum equatorial streamer observed with UVCS in 2008, at the same coronal height \citet[][]{dolei2015} find a much higher hydrogen outflow speed, about 140~km~s$^{-1}$.
In general, these studies indicate that O$^{5+}$ ions flow outward at a higher speed than protons.
This difference of flow speed suggests a different energy deposition rate in the solar wind acceleration region for the H$^{0}$ and O$^{5+}$ components, not only in the polar coronal holes, where this phenomenon has been examined in depth \citep[e.g.,][]{cranmer2017} but also in the proximity of the coronal/heliospheric current sheet.

The present data, extending previous results out to 6~$R_\odot$, indicate that the neutral hydrogen/proton component in proximity of the current sheet reaches a speed of $\approx160$~km~s$^{-1}$ at 4~$R_\odot$ and this maintains a fairly constant value out to 6~$R_\odot$.
It is interesting to note that if the hydrogen kinetic temperature is assumed to be isotropic in the computation of Doppler dimming, the wind speed derived from Metis data is higher by $\approx40$~km~s$^{-1}$, but still lower than the value, 260~km~s$^{-1}$ (with large uncertainties as reported by the authors), measured during a balloon experiment along a streamer by \citet{gopalswamy2021} in 2019.

In this study, we find that the slowest wind is confined within a narrow layer of enhanced plasma density, $20\degr$ wide, embedding the current sheet out to 6~$R_\odot$ in the solar corona.
According to Ulysses \emph{in situ} measurements along its out-of-ecliptic journey, during solar minimum the slow solar wind was found to be restricted to a narrow region around the equator about $13\degr$ wide, if continuously detected throughout a solar rotation \citep[][]{woch1997}.
The existence of a sharper interface boundary between the fast and slow wind in the heliosphere was also supported by other parameters such as composition measurements \citep[][]{vonsteiger2000}.
Thus, the width of the layer confining the slowest wind is probably overestimated because of projection effects, or furtherly reduced in the transition to the heliosphere.
The \emph{in situ} instruments of Solar Orbiter will be able to cast light on this issue when the angle of the orbital plane of the spacecraft will increase relative to the ecliptic one.

The solar wind flowing at velocity of about 160~km~s$^{-1}$ in the observed coronal region from 4 to 6~$R_\odot$ is still quite slower than the wind detected in the heliosphere.
Under the assumption that the source region of the solar wind observed with Metis on May~15, 2020, at about 12:00~UT is steady, it takes almost 6~days to face Earth, and, in turn, the spacecraft orbiting at L1.
This time interval accounts for the solar rotation, the different heliographic longitudes of Solar Orbiter and Earth, and the fact that the plasma observed by Metis is outflowing at about $90\degr$ with respect to the Sun-Solar Orbiter direction.
Wind velocity data collected by the Solar Wind Experiment \citep[][]{ogilvie1995} at L1 are then ballistically mapped back to 6~$R_\odot$.
Hence, the plasma outflows observed with Metis should be further accelerated to the final speed of 300~km~s$^{-1}$, measured locally at L1 about 11.5~days later.
This confirms that the main component of the solar wind, formed by protons, undergoes significant acceleration processes beyond the coronal heights considered in this study.

The physical processes underlying the acceleration of the solar wind plasma to the supersonic speeds observed in the outer corona may reside in the topology of the magnetic field and the flux-tube expansion rate, that determines how rapidly the field strength falls off with distance, and thus the radial range over which the energy -- assumed to depend on the local field strength -- is deposited.
Slow wind conditions are obtained when the field falls off very rapidly and the heating is concentrated near the coronal base below the sonic point, with the effect of an increase of the mass flux density but a decrease of the energy available per proton \citep[e.g.,][]{leer1980}.
In particular, the flux-tube expansion near the helmet streamer cusp is non-monotonic, that is, the open flux tubes just inside the boundary of the coronal hole expand very rapidly near the cusp, and beyond the cusp the flux tubes reconverge \citep[see][]{wang1994,noci1997}.
As a consequence of this magnetic configuration, the flow speed remains low at least up to the streamer cusp.
Along the equatorial boundary of the coronal hole, the adjacent coronal-hole wind may act to accelerate the plasma sheet wind via shear flow interactions.
In the region of interface between the fast and slow wind, from the sonic point to the Alfv\'en surface (i.e., between about 2 and 11~$R_\odot$), the shear-driven Kelvin-Helmholtz (KH) instability should arise.
Although the magnetic field tends to stabilize the KH instability, the streamer belt region may not have sufficiently ordered fields and it is possible that the compressible turbulence would be excited.
In these regions, the dissipation of turbulence might lead to greater heating and an acceleration of the plasma flow to higher velocities would be expected.

\section{Conclusions}

The analysis of the first Metis observations, obtained on May~15, 2020, has allowed us to produce velocity maps of the outer solar corona and identify the plasma sheet, about $\pm10\degr$ wide, embedding the current sheet.
This equatorial layer of enhanced density is the site of the slowest wind component with almost uniform speed both in latitude, within $\pm0.6~R_\odot$, and along the radial direction out to 6~$R_\odot$.
This slow wind is flowing at $160\pm18$~km~s$^{-1}$, under the assumption that the kinetic temperature of the neutral hydrogen/protons is anisotropic relative to the direction of the magnetic field.
In the case of an assumption of isotropy of the kinetic temperature of the hydrogen atoms along and perpendicularly to the magnetic field, the outflow velocity is about 40~km~s$^{-1}$ higher.
The speed values of the slowest component of the coronal wind indicate that the wind plasma, as expected, has to undergo further significant acceleration processes beyond 6~$R_\odot$ in order to reach its interplanetary value of 300~km~s$^{-1}$ measured at L1.
Outside the boundaries of the equatorial layer of enhanced density plasma, the wind velocity rapidly increases.
The observed high velocity gradient in latitude is marking the transition between slow and fast wind in the corona.
This latitudinal transition, from the plasma expansion rate pertaining to the streamer to that typical of the polar holes, is at present under further study.

\begin{acknowledgements}
This work is dedicated to the memory of Giancarlo Noci.
\\
Solar Orbiter is a space mission of international collaboration between ESA and NASA, operated by ESA.
\\
The Metis programme is supported by the Italian Space Agency (ASI) under the contracts to the co-financing National Institute of Astrophysics (INAF): Accordi ASI-INAF N. I-043-10-0 and Addendum N. I-013-12-0/1, Accordo ASI-INAF N.2018-30-HH.0, and under the contracts to the industrial partners OHB Italia SpA, Thales Alenia Space Italia SpA and ALTEC: ASI-TASI N. I-037-11-0 and ASI-ATI N. 2013-057-I.0.
\\
Metis was built with hardware contributions from Germany (Bundesministerium f\"ur Wirtschaft und Energie (BMWi) through the Deutsches Zentrum f\"ur Luft- und Raumfahrt e.V. (DLR)), from the Academy of Science of the Czech Republic (PRODEX) and from ESA.
\\
The authors acknowledge the unique contribution of Dr.
Bogdan Nicula from ROB (Belgium): his continuous support and willingness have been a fundamental key to generate images with JHelioviewer.
\\
Wilcox Observatory data used in this study was obtained via the web site \url{http://wso.stanford.edu}, courtesy of J.T. Hoeksema. The Observatory is currently supported by NASA.
\end{acknowledgements}

\bibliographystyle{aa}
\bibliography{biblio}
\end{document}